\documentclass[showpacs,amsmath,amssymb,amsfonts,a4paper,aps,prd]{revtex4}
\usepackage[colorlinks, linkcolor={RoyalBlue},citecolor={Red}]{hyperref}
\usepackage{graphicx,color}
\usepackage{float}
\usepackage[utf8]{inputenc}
\usepackage{multirow}

\usepackage[dvipsnames]{xcolor}

\begin{document}

\title{ General Scalar-Tensor cosmology: Analytical solutions via Noether symmetry }

\author{Erfan Massaeli$^1$}
\email{erfan.massaeli@gmail.com}
\author{Meysam Motaharfar$^1$}
\email{mmotaharfar2000@gmail.com}
\author{Hamid Reza Sepangi$^1$}
\email{hr-sepangi@sbu.ac.ir}
\affiliation{$^1$Department of Physics, Shahid Beheshti University, G. C., Evin,Tehran 19839, Iran}

\begin{abstract}
We analyze the cosmology of a general Scalar-Tensor theory which encompasses generalized Brans-Dicke theory, Gauss-Bonnet gravity, non-minimal derivative gravity, generalized Galileon gravity and also the general k-essence type models. Instead of taking into account phenomenological considerations we adopt a Noether symmetry approach, as a physical criterion, to single out the form of undetermined functions in the action. These specified functions symmetrize equations of motion in the simplest possible form which result in exact solutions. Demanding de Sitter,  power-law and bouncing universe solutions in the absence and presence of matter density leads to exploring new as well as well-investigated models. We show that there are models for which dynamics of the system allow transition from a decelerating phase (matter dominated era) to an accelerating phase (dark energy epoch) and could also lead to general Brans-Dicke with string correction without a self-interaction potential. Furthermore, we classify the models  based on phantom or quintessence dark energy point of view. Finally, we obtain the condition for stability of a de Sitter solution for which the solution is an attractor of the system.
\end{abstract}

\date{\today}

\maketitle

                                                                                    \section{introduction}

The investigation of alternative theories of gravity chiefly arises from open issues in cosmology, astrophysics, quantum field theory and Mach's Principle. In fact, long-standing problems like the initial singularity, flatness, horizon and relic amongst others, articulate that the standard model of cosmology  which is based on particle physics standard model and General Relativity (GR) fails when one wants to portray the universe in its entirety,  particularly when the extreme regimes of ultraviolet scales are concerned. The consequences of these shortcomings and most importantly the absence of an ultimate quantum theory of gravity provides an incentive to consider modifications to GR in order to construct a semi-classical description towards quantization. These theories are aimed at addressing gravitational interactions by adding physically motivated, non-minimally coupled scalar fields or higher-order curvature invariants, like the inclusion of Gauss-Bonnet term in the Einstein-Hilbert action. Therefore, to obtain the low-energy effective action of quantum gravity on scales closer to Planck scale, one needs the inclusion of such corrective terms \cite{corrective term, Capozziello:2011et}.

The enthusiasm in considering such an approach in the cosmology of early universe stems from the fact that Extended Theories of Gravity (ETG) can ``naturally" reproduce inflationary behavior due to the existence of a non-minimal coupled scalar field to curvature, its higher orders and the kinetic term. Therefore, such models are able to overcome the aforementioned shortcomings of the standard model of cosmology \cite{Capozziello:2011et} and seem also capable of justifying several observational data coming from various sources.
In addition, the Mach's Principle which states that a local inertial frame is determined by the average motion of remote astronomical objects, has brought about further incentives to modify GR \cite{local frame}. Consequently, the gravitational coupling can be scale-dependent whereby the concept of inertia and the equivalence principle have to be revised since there is no a priori reason to constrain the gravitational Lagrangian to a linear function of the Ricci scalar $R$, minimally coupled to matter fields \cite{non-linear action}.

In recent years, ETGs are playing an absorbing role in depicting today's observable universe. In fact, the spectacular amount of high quality data produced over the past few decades seem to shed new light on the effective picture of the universe. Type Ia Supernovae (SNeIa) \cite{SNIa}, Large-Scale Structure (LSS) \cite{observations-LSS}, Baryon Acoustic Oscillations (BAO) \cite{BAO}, anisotropies in the Cosmic Microwave Background Radiation (CMBR) \cite{observations-CMB}
and matter power spectrum extracted from the wide and deep galaxy surveys provide incontrovertible evidences whereby the standard model of cosmology  should radically be revised at cosmological scales. Specifically, the ubiquitous  $\Lambda CDM$ Model indicates that the baryon contribution to the total matter-energy budget is roughly around $(\sim \% 4)$, while Cold Dark Matter (CDM) represents the bulk of the clustered large-scale structure by $(\sim \% 25)$ and the so-called cosmological constant $\Lambda$ plays the role of Dark Energy (DE) contributing  $(\sim \% 75)$ \cite{dark-energy} to the total matter-energy supply.

The incentive to search for alternative models of dark energy \cite{alternative models} stems from the fact that the $\Lambda CDM$ model is affected by strong theoretical shortcomings \cite{ de sitter shortcomings} whereas the model is incredibly compatible with a broad range of data \cite{range of data}. The validity of GR at large astrophysical and cosmological scales has never been confirmed but merely assumed \cite{GR}. Nonetheless dark energy models are primarily built on the implicit assumption that Einstein's GR is the correct theory of gravity. Therefore, it is conceivable that both the existing cosmic acceleration and missing relic are nothing else but a signal of a breakdown of GR. In this sense, GR could fail in representing self-consistent pictures both at ultraviolet (early universe) and infrared scales (late universe). Hence, one possible way to explain the current accelerating epoch is to consider scalar-tensor theories where different couplings of the scalar field to curvature exist. \cite{scalar-tensor}.
Scalar-tensor theories have been extensively considered among all possible explanations to describe cosmic acceleration,  since the coupling of the scalar field to curvature naturally appear in the process of quantization on a curved space-time \cite{quantization}, from compactification of higher dimensional gravity theories \cite{compactification} to next to leading order corrections in the $\alpha^{\prime}$ expansion of string theory \cite{alpha expansion} where $\alpha^{\prime}$ is the inverse string tension. From a mathematical point of view, these corrective terms, in fact, lead to the corresponding effective energy-momentum tensor for modified Einstein-Hilbert action which provides an elegant methodology to deal with the eccentric dark energy,  violating energy conditions  in cosmology.

In low-energy effective string theory which is a leading candidate for quantum theory of gravity there is a scalar field degree of freedom $\phi$ known as the dilaton which couples  to the Ricci scalar $R$ in the form $f(\phi) R$ \cite{f(phi)R}. A field coupling of the form $\xi_{1}(\phi) R_{GB}$, where $R_{GB}$ is the Gauss-Bonnet curvature \cite{Gauss-Bonnet}
also arises as a higher order string correction to the low-energy effective string action \cite{alpha expansion}. Furthermore, the higher order string corrections include a non-minimal derivative term  $\xi_{2}(\phi)G^{\mu\nu} (\partial \phi)^{2}$ as well as a non-linear self-interaction of the form $\xi_{3}(\phi) (\partial \phi)^{2} \Box \phi$,  where for constant $\xi_{3}(\phi)$ this term reduces to a covariant Galileon field that respects the Galilean symmetry $\partial_{\mu}\phi \rightarrow \partial_{\mu} \phi + A_{\mu}$ in Minkowski space-time \cite{Galileon}
and indeed a non-canonical term like $\xi_{4}(\phi) (\partial \phi)^{4}$ keeping the field equations second order.
One may also accommodate non-linear field derivative terms such as $\xi_{3}(\phi) X^{n} \Box \phi$, which are usually referred to as  generalized Galileons  and $\xi_{4}(\phi) X^{m}$ which are known as k-essence, in order to have a further general action where for $n=1$ and $m=2$  reduces to that of the general Brans-Dicke with string corrections. Since the existence of each corrective term  accentuates a fresh characteristic of the model involved and that in turn points to a somewhat new description of various phenomena, numerous works have been carried out around such  combinations  with different couplings  between them, hoping to eventually produce a cohesive description of the  observed data, for a comprehensive review on Scalar-Tensor inflationary and dark energy models see \cite{inflation-review} and  \cite{ dark-energy-review}.
Beside compatibility with  observational data, an important criteria that a Scalar-Tensor theory must satisfy in order to be viable is to reproduce the desired dynamics of the universe including an inflationary era, followed by a radiation  and matter dominated era and finally current accelerating epoch in which the theory must possess a future (or at least a meta-stable) de Sitter asymptote which is indispensable to portray existing dark energy \cite{minimal criteria}.  Therefore, any viable model should be able  to satisfy the above properties.

Symmetries have always been playing a principal role in the conceptual discussions of classical and quantum physics. The chief reason is that various conservation laws, namely energy, momentum, angular momentum and so forth provide the integrals of motion for a given dynamical system due to the existence of some type of symmetry in that system. From a further general perspective, it can be demonstrated that such laws of conservation are particular cases of the so-called Noether theorem pursuant to which a first integral of motion is resulted for every one-parameter group of coordinate transformation on the configuration space of a system keeping the Lagrangian invariant \cite{first integral}. In mathematical language, this means that if the vector field $X$ is the generator of the above diffeomorphism, the Lie derivative of the Lagrangian along $X$ should vanish; $\mathcal{L}_{X}L=0$ \cite{Noether condition}. The idea of using Noether symmetry in cosmology is not new and numerous works have indeed been done in the literature along this line. In this context, the first use of Noether symmetry as a selection criteria in scalar-tensor theory, since such theories usually include unspecified functions which would increase arbitrariness, was exploited in \cite{Capozziello}. For a review on Noether symmetry see \cite{Noether-symmetry}.

Having the above points in mind, the main goal of this paper is to explore Noether symmetry in general scalar-tensor theories which allow us to find analytical solutions for the variety of models and also provide cosmologically viable models satisfying minimal criteria. The layout of the paper is the following. In section II, the action for the model is presented, followed by the corresponding point-like Lagrangian and equations of motion adopting FRLW metric. Section III is devoted to discussing Noether theorem in general and the way it reduces the dynamics of the system and brings the possibility of finding exact solutions. In what follows, we attempt to solve a coupled partial differential equation in order to find the functional form of the undetermined functions. Inserting these functions in two Friedmann equations, we demand for exact de Sitter, power law and bouncing universe solutions which lead to explore novel and well-investigated models. Therefore, we present models which are able to transit from decelerating phase (matter dominated era) into accelerating epoch (de Sitter phase) and also models which describe a sequence from the inflationary era to accelerating phase at late times. In section IV, we  find the condition for stability of the de Sitter solution through perturbing the action up to second order around a de Sitter background. Finally, conclusions are drawn in the last section. Throughout the paper we adopt the Planckian units $8 \pi G = \hbar = c =1$ and the metric signature $(+ , -,  -,  -)$.

                                                                    \section{The Model and background equations}
We start with the following action

\begin{align}\label{aa1}
 S &= \int d^{4} x \sqrt{-g} \Big[\frac{1}{2}f(\phi) R + P(\phi, X) - \xi_{1}(\phi)R_{GB} - \xi_{2}(\phi) G^{\mu\nu}\partial_{\mu} \phi \partial_{\nu} \phi  - G(\phi, X) \Box \phi\Big]+ L_{m},
\end{align}
where $g$ is determinant of the metric in four dimensions, $R$ is the Ricci scalar,  $R_{GB} = R^{2} - 4 R^{\mu\nu}R_{\mu\nu}+R^{\rho \sigma \mu\nu}R_{\rho \sigma \mu \nu}$  is the Gauss-Bonnet term in which $R_{\mu\nu}$, $R_{\rho \sigma \mu \nu}$ and $G_{\mu\nu} = R_{\mu\nu} - \frac{1}{2} g_{\mu\nu} R$  are the Ricci, Riemann and Einstein tensors respectively. The scalar field is represented by  $\phi$ with $X = - \frac{1}{2}g^{\mu\nu} \partial_{\mu} \phi \partial_{\nu} \phi$ being the kinetic term and $L_{m}$ the matter Lagrangian. To obtain the corresponding point-like Lagrangian we consider
\begin{align}
P(\phi, X) = \omega(\phi)X - V(\phi) - \xi_{4}(\phi) X^{m}, \ \ \ \ \ \ \  G(\phi, X) =  \xi_{3}(\phi) X^{n}.
\end{align}
Where $n$ and $m$ are positive constant. In fact such an action for $\xi_{2}(\phi)=0$ reduces to that for which non-Gaussianities have been investigated in \cite{second order action} and also  to a general Brans-Dicke with string corrections for $n=1$ and $m=2$, whose cosmological perturbations were  studied in \cite{alpha expansion}. Let us proceed further by considering the spatially flat FLRW metric. The corresponding point-like Lagrangian for action (\ref{aa1}) is then given by
\begin{align}\label{a1}
\nonumber L(a, \dot a, \phi, \dot \phi) &= - 3 a \dot a ^{2} f(\phi) - 3 a^{2} \dot a \dot \phi f^{\prime}(\phi) + \frac{1}{2} a^{3} \dot \phi^{2} \omega(\phi) + 8 \dot a^{3} \dot \phi \xi^{\prime}_{1}(\phi)  - \frac{3}{2} a \dot a^{2} \dot \phi^{2} \xi_{2}(\phi) + \frac{6n}{2n+1} a^{2} \dot a \dot \phi X^{n}\xi_{3}(\phi) \nonumber \\&- \frac{2}{2n+1} a^{3} X^{n+1} \xi^{\prime}_{3}(\phi) - a^{3} X^{m} \xi_{4}(\phi) - a^{3} V(\phi) - \rho_{0m},
\end{align}
with $a(t)$ being the scale factor as a function of cosmic time,  $L_{m} =- \rho_{0m} a^{-3}$ in which $\rho_{0m}$ is an integration constant associated with the matter content, $X = \frac{1}{2} \dot \phi^{2}$ and a dot represents derivative with respect to cosmic time whereas a prime denotes derivative with respect to the scalar field. One can also find the following zero energy condition (Hamiltonian constraint) associated with  Lagrangian (\ref{a1})

\begin{align}
 E_{L}& \equiv3 H^{2} f + 3 H \dot f - 24 H^{3} \dot \xi_{1} + \frac{9}{2} H^{2} \dot \phi^{2} \xi_{2} - 6 n H \dot \phi X^{n} \xi_{3} + X^{n} \dot \phi \dot \xi_{3} + (2m-1) X^{m} \xi_{4}- \omega X - V(\phi) - \rho_{m} = 0, \label{w1}
\end{align}
where $H = \frac{\dot a}{a}$  is the Hubble parameter.  Eq. (\ref{a1}) is also known as the first modified Friedmann equation which corresponds to the $G_{00}$ component. Furthermore, the equations of motion can be obtained by varying the Lagrangian (\ref{a1}) with respect to $a$ and $\phi$, respectively
\begin{align}
E_{a} &\equiv\nonumber 3 H^{2}f + 2 \dot H f + 2 H \dot f + \ddot f - 16 H \dot H \dot \xi_{1}- 16 H^{3} \dot \xi_{1} \nonumber  - 8 H^{2} \ddot\xi_{1}+ \frac{1}{2}(3 H^{2} + 2 \dot H) \dot \phi^{2} \xi_{2} + H (\dot \phi^{2} \dot \xi_{2} + 2 \dot\phi \ddot\phi \xi_{2})\\& -  n X^{n-1} \dot \phi^{2} \ddot \phi \xi_{3} - X^{n} \dot \phi \dot \xi_{3} - X^{m}\xi_{4} + \omega X - V= 0, \label{w2}\\E_{\phi}& \equiv\nonumber  \Big(\omega - 3 H^{2} \xi_{2} +  6 n^{2} H \dot \phi X^{n-1} \xi_{3}- 2(n+1) X^{n}\xi^{\prime}_{3} \nonumber  - m(2m-1) X^{m-1} \xi_{4}  \Big) \ddot \phi + \Big(3 H \omega + \omega^{\prime} \dot \phi - 6 H \dot H \xi_{2}- \frac{3}{2} H^{2} \dot \phi \xi_{2}^{\prime} \\&\nonumber  \nonumber  - 9 H^{3} \xi_{2} +  9 n H^{2} \dot \phi X^{n-1} \xi_{3} + 3 n \dot H \dot \phi X^{n-1} \xi_{3} \nonumber+ 3n  H^{2} \dot \phi^{2} X^{n-1} \xi_{3}^{\prime} -  6 H X^{n} \xi_{3}^{\prime} -  X^{n} \dot \phi \xi_{3}^{\prime \prime} - m \dot \phi X^{m-1} \xi_{4}^{\prime} - 3 m H X^{m-1} \xi_{4}\Big) \dot \phi  \\& - 6 H^{2} f^{\prime}-  3 \dot H f^{\prime} - \omega^{\prime} X + 24 H^{4} \xi_{1}^{\prime} + 24 H^{2} \dot H \xi_{1}^{\prime}+ X^{m} \xi_{4}^{\prime} + V^{\prime} = 0. \label{w3}
\end{align}
Here, $E_{a}$ is the second modified Friedmann equation corresponding to the $G_{ii}$ components of Einstein field equations and $E_{\phi}$ is the modified Klein-Gordon equation. We note that because of the existence of Bianchi identity, $\dot \phi E_{\phi} + \dot E_{L} + 3 H (E_{L}-E_{a})=0$, only two of the above equations are independent.
\section{Noether symmetry approach}

Generally speaking, the Noether symmetry plays a vital role in physics since it can be used to simplify a given system of differential equations as well as to determine the integrability of the system. As we mentioned in the introduction, the associated Noether conserved charges in Scalar-Tensor theories typically reduce dynamics of the system in such a way as to result in determining the unspecified functions in the action and hence extract exact cosmological solutions. As is well known,  Noether symmetry exists for the Lagrangian $L(q_{i}, \dot q_{i})$ if
\begin{align}\label{ss1}
\mathcal{L}_{X} L= X L(q_{i}, \dot q_{i}) = 0.
\end{align}
Here,  $\mathcal{L}_{X}$ is the Lie derivative with respect to Noether vector $X$ which is defined on the tangent space $\mathcal{T}\mathcal{Q}=\{q_{i}, \dot q_{i}\}$. The immediate consequence of the above condition for Lagrangian $L(q_{i}, \dot q_{i})$ can be expressed as the Cartan-one-form as follows

\begin{align}
i_{X} \theta_{L}  = \Sigma_{0},
\end{align}
and

\begin{align}
\theta_{L} = \frac{\partial L}{\partial \dot q_{i}} d q^{i},
\end{align}
where $i_{X}$ is the interior derivative and $\Sigma_{0}$ represents the conserved quantity. In fact, the existence of Noether symmetry is connected to the existence of a vector field which is expressed according to

\begin{align}
X = \alpha^{i}(q_{i}) \frac{\partial}{\partial q_{i}} + \frac{ d \alpha^{i}(q_{i})}{d t} \frac{\partial}{\partial \dot q_{i}}.
\end{align}
Therefore, the existence of a conserved quantity assures the existence of a cyclic variable under the point transformation

\begin{align}
i_{X} d Q^{1} = 1, \ \ \ \ \ i_{X} d Q^{i} = 0, \ \ \ \ i \neq 1,
\end{align}
where $Q^{1}$ is a cyclic coordinate which leads to the conserved quantity $\Sigma_{0}$ in a new coordinate system. In other words, the existence of cyclic variables means that the dynamics of the system is simplified and allows one to integrate equations of motion. This is to say that the conserved quantities are revealed ``directly'' through the dynamics or alternatively, it is the dynamics that give rise to conserved quantities. In fact, Noether symmetry selects the dynamics and results in conserved quantities where the dynamical equations are simplified for some particular functional forms of the undetermined functions. In addition, it should also be mentioned that in terms of cyclic variables the conserved quantities are nothing but the corresponding conjugate momenta. In fact, the momenta will be conserved for the selected functional forms by Noether symmetry. In our case, upon having a look at Lagrangian (\ref{a1}), one finds that although the action has a complicated form and contains several terms, it only has a two dimensional phase space $\mathcal{Q}\{a, \phi\}$.

Let us now define the vector field

\begin{align}
X = A(a, \phi) \frac{\partial }{\partial a}+ B(a, \phi) \frac{\partial }{\partial \phi}+\dot A(a, \phi) \frac{\partial }{\partial \dot a}+\dot B(a, \phi) \frac{\partial }{\partial \dot a},
\end{align}
where

\begin{align}
\dot A(a, \phi) = \dot a \frac{\partial A}{\partial a} + \dot \phi  \frac{\partial A}{\partial \phi},  \ \ \ \   \dot B(a, \phi) = \dot a \frac{\partial B}{\partial a} + \dot \phi  \frac{\partial B}{\partial \phi},
\end{align}
for which  condition $(\ref{ss1})$ for Lagrangian (\ref{a1}) becomes

\begin{align}
\mathcal{L}_{X}L(a, \dot a, \phi, \dot \phi)= X L = 0.
\end{align}
The above equation can be expanded as follows

\begin{widetext}
\begin{align}\label{ss2}
\nonumber &\left(-3 a B f^{\prime} - 6 a A_{,a}f  - 3 a^{2} B_{,a} f^{\prime} - 3 A f\right) \dot a^{2} + \left(\frac{3}{2} a^{2} A  \omega +  \frac{1}{2} a^{3} B \omega^{\prime} - 3 a^{2} A_{,\phi} f^{\prime}+  a^{3} B_{,\phi} \omega \right) \dot \phi^{2}+ \left( - 3 a^{2} A_{,a} f^{\prime} - 6 a A_{,\phi} f \right. \\& \left.+ a^{3} B_{,a} \omega  - 3 a^{2} B_{,\phi} f^{\prime} - 6 a A f^{\prime} - 3 a^{2} B f^{\prime \prime}\right) \dot a \dot \phi+ \left(- \frac{3}{2} A \xi_{2} + 24 A_{,\phi} \xi_{1}^{\prime} - \frac{3}{2} a B \xi_{2}^{\prime} - 3 a A_{,a} \xi_{2} - 3 a B_{,\phi} \xi_{2}\right) \dot a^{2} \dot \phi^{2}\nonumber + \left(8 B \xi_{1}^{\prime \prime} \right.\\&\left.  \nonumber+ 24 A_{,a} \xi_{1}^{\prime} + 8 B_{,\phi} \xi_{1}^{\prime} - 3 a B_{,a} \xi_{2}\right) \dot a^{3} \dot \phi - 3 a A_{,\phi} \xi_{2} \dot a \dot \phi^{3}+ 8 B_{,a} \xi_{1}^{\prime} \dot a^{4} + \left( - \frac{2n+2}{2n+1} a^{3} B_{,a} \xi_{3}^{\prime} + \frac{12n}{2n+1} a A \xi_{3} + \frac{6n}{2n+1} a^{2} B \xi_{3}^{\prime}\right.\\&\left.\nonumber + \frac{6n}{2n+1} a^{2} A_{,a} \xi_{3}+6 n a^{2} B_{,\phi} \xi_{3} \right) \dot a \dot \phi X^{n} + 6n a^{2} B_{,a}\xi_{3} \dot a^{2} X^{n} - m a^{3} B_{,a}\xi_{4} \dot a \dot \phi X^{m-1}+  \left(\frac{6n}{2n+1} a^{2}A_{,\phi} \xi_{3}- \frac{3}{2n+1} a^{2} A \xi_{3}^{\prime} \right.\\&\left. - \frac{2n+2}{2n+1} a^{3} B_{,\phi} \xi_{3}^{\prime} - \frac{1}{2n+1} a^{3} B \xi_{3}^{\prime \prime}\right)X^{n+1}- \left(a^{3} B \xi_{4}^{\prime} + 3 a^{2} A \xi_{4} + 2 m a^{3} B_{,\phi} \xi_{4}\right)X^{m} - a^{3} B V^{\prime}- 3 a^{2} A V = 0,
\end{align}
\end{widetext}
where the comma denotes partial derivative. The coefficients of the above equation take various forms for  $m=n+1$, $m \neq n+1$ and $n=1$ hence there are several cases that should be investigated.

\subsection{Case $m \neq n+1$}

Eq. (\ref{ss2}) is a polynomial in the terms of $\dot a^{2},\dot \phi^{2},\dot a \dot \phi, \dot a^{2} \dot \phi^{2}, \dot a^{3} \dot \phi, \dot \phi^{3} \dot a, \dot a^{4}, \dot a \dot \phi X^{n}, \dot a^{2} X^{n}, \dot a \dot \phi X^{m-1}, X^{n+1}$, $X^{m}$ and terms which contain any time derivatives of configuration space variables and for it to be zero, each coefficient should vanish separately. Therefore, Eq. (\ref{ss2}) leads to a system of coupled partial differential equations as follows

\begin{align}
-&3 a B f^{\prime} - 6 a A_{,a}f  - 3 a^{2} B_{,a} f^{\prime} - 3 A f=0, \\
+&\frac{3}{2} a^{2} A  \omega +  \frac{1}{2} a^{3} B \omega^{\prime} - 3 a^{2} A_{,\phi} f^{\prime}+  a^{3} B_{,\phi} \omega=0,\\
 -& 3 a^{2} A_{,a} f^{\prime} - 6 a A_{,\phi} f + a^{3} B_{,a} \omega - 3 a^{2} B_{,\phi} f^{\prime} - 6 a A f^{\prime} -3 a^{2} B f^{\prime \prime}=0, \\
- &\frac{3}{2} A \xi_{2} + 24 A_{,\phi} \xi_{1}^{\prime} - \frac{3}{2} a B \xi_{2}^{\prime} - 3 a A_{,a} \xi_{2} - 3 a B_{,\phi} \xi_{2}=0,\\
+&8 B \xi_{1}^{\prime \prime} + 24 A_{,a} \xi_{1}^{\prime} + 8 B_{,\phi} \xi_{1}^{\prime} - 3 a B_{,a} \xi_{2}=0,\\
+&6 n (2n+1) a^{2} B_{,\phi} \xi_{3} - (2n+2) a^{3} B_{,a} \xi_{3}^{\prime} + 12 n a A \xi_{3}+ 6n a^{2} B \xi_{3}^{\prime}+ 6n a^{2} A_{,a} \xi_{3} =0,\\
+&6na^{2}A_{,\phi} \xi_{3} - (2n+2) a^{3} B_{,\phi} \xi_{3}^{\prime} - 3 a^{2} A \xi_{3}^{\prime} -  a^{3} B \xi_{3}^{\prime \prime}=0,\\
+&a^{3} B \xi_{4}^{\prime} + 3 a^{2} A \xi_{4} + 2 m  a^{3} B_{,\phi} \xi_{4} = 0,\\
-& a^{3} B V^{\prime} - 3 a^{2} A V = 0, \\
& 8 B_{,a} \xi_{1}^{\prime} = 0,  \ \ \ 3 a A_{,\phi} \xi_{2} = 0, \ \ \ 6n a^{2} B_{,a} \xi_{3}=0, \label{af1}  \ \ \ \ m a^{3} B_{,a} \xi_{4} = 0 ,
\end{align}
where $n\neq - \frac{1}{2}$. In the case $\xi_{1}^{\prime}, \xi_{2}, \xi_{3}, \xi_{4}\neq 0$, Eqs. (\ref{af1}) result in

\begin{align}\label{a2}
A(a, \phi)= A(a), \ \ \ \ \ \ \ \ \ B(a,\phi) = B(\phi).
\end{align}
At this point it should be emphasized that for the case $\xi_{i}(\phi)=0$ where $i$ runs from 1 to 4 and $\omega(\phi) \sim 1$, the action reduces to the non-minimal coupling scalar-tensor theory in which one can find more general forms for $A(a,\phi)$ and $B(a, \phi)$ since the condition (\ref{af1}) will no longer exist  \cite{nonminimal Noether}. In fact, the Noether vector in our case is different because of the existence of $\xi_{i}(\phi)$ and $\omega(\phi)$ which makes it a subclass of the Noether vector in \cite{nonminimal Noether} in the absence of $\xi_{i}(\phi)$ and $\omega(\phi)$. Therefore, one could not expect our results to coincide with the  results  found in \cite{nonminimal Noether} in the absence of $\xi_{i}(\phi)$ and $\omega(\phi)$. Using (\ref{a2}), the partial differential equations will change into a coupled system of differential equations

\begin{align}
-&3 a B f^{\prime} - 6 a A_{,a}f - 3 A f=0 \label{a3},\\
+&\frac{3}{2} a^{2} A  \omega +  \frac{1}{2} a^{3} B \omega^{\prime}+  a^{3} B_{,\phi} \omega=0,\label{a4}\\
-& 3 a^{2} A_{,a} f^{\prime} - 3 a^{2} B_{,\phi} f^{\prime} - 6 a A f^{\prime} - 3 a^{2} B f^{\prime \prime}=0,\label{a5}\\
- &\frac{3}{2} A \xi_{2} - \frac{3}{2} a B \xi_{2}^{\prime} - 3 a A_{,a} \xi_{2} - 3 a B_{,\phi} \xi_{2}=0,\label{a6}\\
+&8 B \xi_{1}^{\prime \prime} + 24 A_{,a} \xi_{1}^{\prime} + 8 B_{,\phi} \xi_{1}^{\prime} =0,\label{a7}\\
 +&6 n (2n+1) a^{2} B_{,\phi} \xi_{3} + 12 n a A \xi_{3} + 6n a^{2} B \xi_{3}^{\prime}+ 6n a^{2} A_{,a} \xi_{3} =0,\label{a8}\\
-& (2n+2) a^{3} B_{,\phi} \xi_{3}^{\prime} - 3 a^{2} A \xi_{3}^{\prime} -  a^{3} B \xi_{3}^{\prime \prime}=0,\label{a9}\\
+&a^{3} B \xi_{4}^{\prime} + 3 a^{2} A \xi_{4} + 2 m  a^{3} B_{,\phi} \xi_{4} = 0,\label{a10}\\
-& a^{3} B V^{\prime} - 3 a^{2} A V = 0.\label{a11}
\end{align}
Condition (\ref{a2}) completely decouples the above system of differential equations and since  $A(a)$ is a function of $a$ only, these equations have a general solution when $A(a) \propto a$. Taking $A= - \frac{q}{3} a$ in Eq. (\ref{a3}), one finds

\begin{align}\label{a12}
f(\phi) = \lambda_{1} \exp \left( \int \frac{q}{B(\phi)} d \phi\right),
\end{align}
where $B(\phi)$ is an arbitrary function of $\phi$. Using Eqs. (\ref{a4} - \ref{a7}, \ref{a10}, \ref{a11}), one can easily integrate them in terms of $B(\phi)$

\begin{align}
 \omega(\phi) &=  \frac{\lambda_{2}}{\left[B(\phi)\right]^{2}} \exp \left( \int \frac{q}{B(\phi)} d \phi\right), \ \ \ \ \ \ \  \xi_{1}(\phi) = \lambda_{3} \int \frac{1}{B(\phi)} \exp \left( \int \frac{q}{B(\phi)} d \phi\right) d \phi \label{sp5} ,\\ \xi_{2}(\phi) &=  \frac{\lambda_{4}}{\left[B(\phi)\right]^{2}} \exp \left( \int \frac{q}{B(\phi)} d \phi\right), \ \ \ \ \ \ \ \xi_{4}(\phi) =   \frac{\lambda_{6}}{\left[B(\phi)\right]^{2m}} \exp \left( \int \frac{q}{B(\phi)} d \phi\right)\label{h1}, \\  V(\phi) &= \lambda_{7} \exp \left( \int \frac{q}{B(\phi)} d \phi\right).\label{sp6}
\end{align}

So far, we have noted that the form of unspecified functions in the action are determined by the functionality of $B(\phi)$.  However, using Eqs. (\ref{a8}, \ref{a9}), we have

\begin{align}\label{a14}
\frac{\xi_{3}^{\prime}}{\xi_{3}} = \frac{-(2n+1) B^{\prime}+ q}{B},  \ \ \ \ \frac{\xi_{3}^{\prime \prime}}{\xi_{3}^{\prime}}=\frac{-(2 n+2) B^{\prime}+q}{B},
\end{align}
from which we see that for a constant $B$ we have an exponential form for the solution and for $B\sim \phi$ the solution will be of the power law form.
Therefore, the form of the undetermined functions in the action can be obtained by the above conditions. The choice $B(\phi) = \mbox{const.} = 1$  in Eqs. (\ref{sp5} - \ref{sp6}) gives

\begin{align}\label{sw1}
\nonumber f(\phi) &= \lambda_{1} \exp (q \phi), \ \ \ \ \omega(\phi) = \lambda_{2} \exp (q \phi), \ \ \ \ \xi_{1}(\phi) = \frac{\lambda_{3}}{q} \exp (q \phi), \ \ \ \ \xi_{2}(\phi) = \lambda_{4}\exp (q \phi),\ \ \ \ \xi_{3}(\phi) = \lambda_{5}\exp (q \phi),\\ \xi_{4}(\phi)& = \lambda_{6}\exp (q \phi), \ \ \ \ V(\phi) = \lambda_{7} \exp(q \phi),
\end{align}
where $\lambda_{i}$, $i$ running from 1 to 7,  are integration constants. In fact, Noether symmetry exists for exponential functions known as dilatonic fields in string theory. In addition,  for $B(\phi) = \phi$ one finds power-law solutions as follows

\begin{align}\label{a17}
\nonumber f(\phi) &=\lambda_{1} \phi^{q}, \ \ \ \ \omega(\phi) = \lambda_{2} \phi^{q-2}, \ \ \ \   \xi_{1}(\phi) = \frac{\lambda_{3}}{q} \phi^{q},\ \ \ \  \xi_{2}(\phi) = \lambda_{4} \phi^{q-2},  \ \ \ \ \xi_{3}(\phi) = \lambda_{5} \phi^{q-2n-1},  \ \ \xi_{4}(\phi) = \lambda_{6} \phi^{q-2m}, \\ V(\phi) &= \lambda_{7} \phi^{q}.
\end{align}
where for $q=1$ and $n=1$, $f(\phi) \sim \phi$, $\omega(\phi) \sim \frac{1}{\phi}$ and $\xi_{3}(\phi) \sim \frac{1}{\phi^{2}}$ which imply that Noether symmetry exists for the so-called coupled Brans-Dicke with Galileon term. Also,  for $q=2$  Noether symmetry exists for non-minimal coupled scalar-tensor $f(\phi) \sim \phi^{2}$ with quadratic potential $V(\phi) \sim \phi^{2}$. In the case $\lambda_{5} = 0$ ($\xi_{3}(\phi)=0$), Eqs. (\ref{a8}, \ref{a9}) will no longer exist whereby condition (\ref{a14}) does not exist and the unknown functions are determined by the functional form of $B(\phi)$ as they were found in (\ref{sp5} - \ref{sp6}).
Such functional forms symmetrize equations of motion which would result in exact solutions. Therefore, upon substituting (\ref{sw1}) in  Eqs. (\ref{w1}, \ref{w2}) instead of using cyclic variables, we can find the solutions since Eqs. (\ref{w1}, \ref{w2}) can be written in a simple form in the following manner

\begin{align}
\nonumber &3 \lambda_{1} \left(H^{2} +  q H \Phi\right) - \frac{1}{2}\lambda_{2} \Phi^{2} - 24 \lambda_{3} H^{3} \Phi + \frac{9}{2} \lambda_{4} H^{2} \Phi^{2} - \left(\frac{1}{2}\right)^{n}\lambda_{5} \left(6 n H \Phi^{2n+1} - q \Phi^{2n+2}\right) + \left(\frac{1}{2}\right)^{m} \lambda_{6} (2m-1) \Phi^{2m} \\&- \lambda_{7} - \rho_{0m} a^{-3} e^{- q \phi} = 0,\label{nb1}\\
\nonumber & \lambda_{1} \left(3 H^{2} + 2 \dot H + 2 q H \Phi  + q^{2} \Phi^{2} +  q \dot \Phi\right) +\frac{1}{2} \lambda_{2} \Phi^{2} -\lambda_{3} \left(16 H \dot H \Phi +16 H^{3} \Phi+ 8 q H^{2}\Phi^{2} +8 H^{2} \dot \Phi\right)   + \frac{1}{2} \lambda_{4} \Phi^{2} \left(3 H^{2} + 2 \dot H\right) \\&+ \lambda_{4} H \Phi \left(q \Phi^{2} + 2\dot \Phi\right)- \left(\frac{1}{2}\right)^{n}\lambda_{5} \left(2n \Phi^{2n}  \dot \Phi+ q \Phi^{2n+2}\right) - \left(\frac{1}{2}\right)^{m} \lambda_{6} \Phi^{2m} - \lambda_{7} = 0,\label{nb2}
\end{align}
where we have divided the whole expression by $\exp (q \phi)$ and defined $\Phi = \dot \phi$ in order to derive above equations. Next, substituting Eqs. (\ref{a17}) in (\ref{w1}, \ref{w2}) one can also find

\begin{align}
\nonumber &3 \lambda_{1} \left(H^{2} +  q H \Phi\right) - \frac{1}{2}\lambda_{2} \Phi^{2} - 24 \lambda_{3} H^{3} \Phi + \frac{9}{2} \lambda_{4} H^{2} \Phi^{2} - \left(\frac{1}{2}\right)^{n}\lambda_{5} \left(6 n H \Phi^{2n+1} - (q-2n-1) \Phi^{2n+2}\right) + \left(\frac{1}{2}\right)^{m} \lambda_{6} (2m-1) \Phi^{2m} \\&- \lambda_{7} - \rho_{0m} a^{-3} \phi^{-q} = 0,\label{cc1}\\
\nonumber & \lambda_{1} \left(3 H^{2} + 2 \dot H + 2 q H \Phi  + q^{2} \Phi^{2} +  q \dot \Phi\right) +\frac{1}{2} \lambda_{2} \Phi^{2} -\lambda_{3} \left(16 H \dot H \Phi +16 H^{3} \Phi+ 8 q H^{2}\Phi^{2} +8 H^{2} \dot \Phi\right)   + \frac{1}{2} \lambda_{4} \Phi^{2} \left(3 H^{2} + 2 \dot H\right) \\&+ \lambda_{4} H \Phi \left(q \Phi^{2} + 2\dot \Phi\right)- \left(\frac{1}{2}\right)^{n}\lambda_{5} \left(2n \Phi^{2n}  \dot \Phi+ (q-1) \Phi^{2n+2}\right) - \left(\frac{1}{2}\right)^{m} \lambda_{6} \Phi^{2m} - \lambda_{7} = 0, \label{cc2}
\end{align}
where again we have divided the whole expression by $\phi^{q}$, assuming $\phi \neq 0$ and defined $\Phi = \frac{\dot \phi}{\phi}$. For the case $\xi_{3}(\phi) = 0$, inserting functional forms (\ref{a12} - \ref{sp6}) into Eqs. (\ref{w1}, \ref{w2}), we find

\begin{align}
 &3 \lambda_{1} \left(H^{2} +  q H \Phi\right) - \frac{1}{2}\lambda_{2} \Phi^{2} - 24 \lambda_{3} H^{3} \Phi + \frac{9}{2} \lambda_{4} H^{2} \Phi^{2} + \left(\frac{1}{2}\right)^{m} \lambda_{6} (2m-1) \Phi^{2m} - \lambda_{7} - \rho_{0m} a^{-3} \exp \left(-\int \frac{q}{B(\phi)}d \phi \right)  = 0,\label{nb3}\\
\nonumber & \lambda_{1} \left(3 H^{2} + 2 \dot H + 2 q H \Phi  + q^{2} \Phi^{2} +  q \dot \Phi\right) +\frac{1}{2} \lambda_{2} \Phi^{2} -\lambda_{3} \left(16 H \dot H \Phi +16 H^{3} \Phi+ 8 q H^{2}\Phi^{2} +8 H^{2} \dot \Phi\right)   + \frac{1}{2} \lambda_{4} \Phi^{2} \left(3 H^{2} + 2 \dot H\right) \\&+ \lambda_{4} H \Phi \left(q \Phi^{2} + 2\dot \Phi\right) - \left(\frac{1}{2}\right)^{m} \lambda_{6} \Phi^{2m} - \lambda_{7} = 0.\label{nb4}
\end{align}
Here we have divided the expression by $\exp \left( \int \frac{q}{B(\phi)} d \phi \right)$ and defined $\Phi = \frac{\dot \phi}{B(\phi)}$ to derive the above equations. Eqs. (\ref{nb1}, \ref{nb2}), (\ref{cc1}, \ref{cc2}) and (\ref{nb3} \ref{nb4}) are our key equations since they can generate exponential, power-law and bouncing universe solutions for each functional form. To this end, one may demand that

$$
H(t)=
\begin{cases}
H_{dS}\rightarrow a(t) = a_{0} e^{H_{dS}t} \ \ \ \ \ $de Sitter$\\
\frac{p}{t}\rightarrow a(t) = a_{0} t^{p} \ \ \ \ \ \ \ \ \ \ \ $power-law$\\
b t\rightarrow a(t) = a_{0} e^{\frac{b}{2}t^{2}} \ \ \ \ \ \ \ $bouncing  universe$
\end{cases}
$$

First, in order to have exact de Sitter solutions, we consider

\begin{align}
H= H_{dS} \rightarrow a(t) =a_{0} e^{H_{dS}t},
\end{align}

$$
\Phi = \Phi_{dS} \rightarrow \phi(t) =
\begin{cases}
\Phi _{ds} t  \  \ \ \ \ \ \ \  \ \ \ \  \ $exponential form$\\
 e^{\Phi_{dS}t}  \ \ \ \ \ \ \ \ \ \ \  \ \ $power-law form$
\end{cases}
$$
and for the case $\lambda_{5}=0$ ($\xi_{3}(\phi)=0$) we have

\begin{align}
\Phi = \Phi_{dS} \rightarrow\int \frac{d \phi}{B(\phi)} = \Phi_{dS} t,
\end{align}
where we have assumed zero constant of integration without loss of generality and also considered both $H_{dS}$ and $\Phi_{dS}$ as constant parameters. Upon substituting $H_{dS}$ and $\Phi_{dS}$, each coupled  Eqs. (\ref{nb1}, \ref{nb2}), (\ref{cc1}, \ref{cc2}) and (\ref{nb3},\ref{nb4}) impose two constraints on the entire free parameters in the action assuming $\rho_{0m}=0$, since there is no matter field in the de Sitter space-time which consequently reduces the degrees of freedom of the parameters of the model by two. In fact, one can write $H_{dS}$ and $\Phi_{dS}$ in terms of the rest of  parameters of the model for each functional form.

In order to find the power-law solution we assume

\begin{align}
H = \frac{p}{t} \rightarrow a(t) = a_{0}t^{p},
\end{align}

$$
\Phi = \frac{z}{t} \rightarrow \phi(t) =
\begin{cases}
z \ln t + \ln \phi_{0}  \  \ \ \ \ \ \ \  \ \ \ \  $exponential form$\\
\phi_{0} t^{z}            \ \ \ \ \ \ \ \ \ \ \ \ \ \ \ \ \ \ \ \ \ \ $power-law form$
\end{cases}
$$
and for $\lambda_{5}=0$ we have $\Phi=\frac{\dot \phi}{B(\phi)}$. Therefore

\begin{align}
\Phi &= \frac{z}{t}\Rightarrow \int \frac{d \phi}{B(\phi)} = z \ln t + \ln \phi_{0}.
\end{align}
Inserting the associated values of $H$ and $\Phi$ for exponential functional forms in Eqs. (\ref{nb1}, \ref{nb2}) with $m=2$, one can find six constraint equations obtained from the coefficients of $\frac{1}{t^{2}}, \frac{1}{t^{4}}$ and $\frac{1}{t^{2n+2}}$ which when solved for $\rho_{0m}=0$,  results in negative values of $n$ which are unphysical from a field theoretical point of view since such negative values of $n$ make the action unrenormalizable and in the presence of matter density leads to

\begin{align}
p =\frac{1}{3}, \ \ \ \ z = \frac{1}{q}, \ \ \ \ n = \frac{1}{2}, \ \ \ \ \lambda_{1} = \frac{3 \rho_{0m}}{5 a_{0}^{3} \phi_{0}^{q}}, \ \ \ \ \lambda_{2} = -\frac{2 \rho_{0m} q^{2}}{5a_{0}^{3} \phi_{0}^{q}}, \ \ \ \ \lambda_{3} = -\frac{27 \lambda_{6}}{16 q^{3}} , \ \ \ \  \lambda_{4} = - \frac{9 \lambda_{6}}{2q^{2}},
\end{align}
where in order to obtain the above solutions we have utilized the following expression

\begin{align}\label{dd1}
\rho_{0m} a^{-3} \exp \left(-\int \frac{q}{B(\phi)}d \phi \right) = \rho_{0m} a_{0}^{-3} \phi_{0}^{- q} t^{-3p - q z },
\end{align}
and have assumed that $B(\phi) = 1$ for the exponential  form. In fact, the solution describes stiff matters with decelerating phase. Moreover, for the case $m \neq 2$ the constraint equations lead to $\lambda_{3} = \lambda_{4} = \lambda_{6} = 0$, indicating that the solution exists for general Branse-Dicke with a Galileon term.

Inserting $H = \frac{p}{t}$ and $\Phi = \frac{z}{t}$ into the first and second Friedmann Eqs. (\ref{cc1}, \ref{cc2}) for $m=2$ ($n \neq 1$) one can find six constraint equations coming from the coefficients of $\frac{1}{t^{2}}, \frac{1}{t^{4}}$ and $\frac{1}{t^{2n+2}}$, while solving them for $\rho_{0m}=0$ results in the following solutions for power law functional forms

\begin{align}
p = 2n+1, \ \ \ \ z = - (6n+2), \ \ \ \ \lambda_{1} = - \frac{2 \lambda_{2}}{3} \left(\frac{3n+1}{2n+1}\right)^{2}, \ \ \ \ \lambda_{3} = - \frac{\lambda_{4}}{4}\left(\frac{3n+1}{2n+1}\right), \ \ \ \  \lambda_{6}= - \frac{1}{2} \lambda_{4} \left(\frac{2n+1}{3n+1}\right)^{2},
\end{align}
where $\omega_{eff} \equiv -1-\frac{2\dot H}{3H^{2}} = -1+ \frac{2}{3 p} = - \frac{6n+1}{6n+3}$. The solution, which is of general Brans-Dicke type with a generalized string correction when accommodating $n$ in the Galileon term, describes quintessence dark energy for $ n>0$ indicating that the universe experiences an accelerating expansion i.e. $\dot a>0$ and $\ddot a>0$. In other words, the model generates repulsive gravitational waves to produce acceleration at cosmological scales for $n>0$.  Quintessence dark energy models have their own particular properties which have extensively been investigated in the past few decades  \cite{phantom and quintessence}. Moreover, the radiation dominated, matter dominated will happen for negative values of $n$ which are unphysical. In addition, associated constraint equations for the case $m \neq 2$ imply $\lambda_{4}=\lambda_{3}=\lambda_{6}=0$ which means that the above solutions exist for general Brans-Dicke with general Galileon term. In fact, we have shown that Noether symmetry exists for both exponential and power-law functional forms in general Brans-Dicke with general Galileon term which leads to a de Sitter solution. This may physically justify the work in \cite{BD sequence} where such  functional forms were assumed for $n=1$ in order to obtain a de Sitter solution. One can indeed obtain the solution in the presence of matter density $\rho_{0m}$ as follows

\begin{align}
p&= \frac{q-2}{3(q-1)}, \ \ \ \ z= \frac{1}{q-1}, \ \ \ \ n=\frac{1}{2}, \ \ \ \ \lambda_{1}= \frac{3 \rho_{0m}(q-1)}{a^{3}_{0}\phi_{0}^{q}(5q-4)}, \ \ \ \ \lambda_{2}= - \frac{2 \rho_{0m} q (q^{2}-1)}{a^{3}_{0}\phi_{0}^{q}(5q-4)},\\
\lambda_{4}&=\frac{8}{3}\lambda_{3}q- \frac{16}{3} \lambda_{3}, \ \ \ \ \ \ \ \ \ \ \ \ \ \ \ \lambda_{6}= \lambda_{3} \left(- \frac{16}{27}q^{3}+ \frac{32}{9} q^{2}- \frac{64}{9} q+ \frac{128}{27}\right),
\end{align}
where we have utilized Eq. (\ref{dd1}) for $B(\phi) =\phi$ in order to obtain the above solutions. The solution has $\omega_{eff}= \frac{q}{q-2}$ which means that it describes quintessence for $\frac{1}{2}<q<1$, phantom dark energy for $1<q<2$, indicating that the universe experiences a contracting expansion (i.e. $\dot a<0$ and $\ddot a>0$, see \cite{phantom and quintessence}), a radiation dominated era for $q=-1$ whereby $\lambda_{2}=0$, and a matter dominated era for $q=0$. In fact, the matter dominated epoch happens for minimal case since $q=0$ results in $f=1$ (assuming $\lambda_{1}=1$). In addition, the universe has decelerating phase for $q<\frac{1}{2}$ and $q>2$.  One may also obtain another power-law solution ($H = \frac{p}{t}$ and $\Phi = \frac{z}{t}$) by inserting $m=2$ in Eqs. (\ref{nb3}, \ref{nb4}). Therefore, utilizing constraint equations emanating from the coefficients of $\frac{1}{t^{2}}$ and  $\frac{1}{t^{4}}$, results in the first set of solution as follows

\begin{align}
p= - \frac{1}{2} q z, \ \ \ \ \lambda_{1}= - \frac{2\lambda_{2}}{3 q^{2}}, \ \ \ \  \lambda_{3} =- \frac{\lambda_{4}}{4 q},  \ \ \ \ \lambda_{6} = - \frac{1}{2} q^{2} \lambda_{4},
\end{align}
where $\lambda_{7}=0$. In this solution $\omega _{eff} =-1 - \frac{4}{3qz}$ which means that the universe goes through an accelerating epoch for $qz<-2$ and $qz>0$ and experiences a decelerating phase for $-2<qz<0$. In other words, the model explains quintessence dark energy for $q z<-2$ and phantom dark energy for $q z>0$. Furthermore,   $z = -\frac{4}{3q}$ leads to a matter dominated era, $z = -\frac{1}{q}$ to a radiation dominated era, $qz \rightarrow \infty$ to de Sitter dark energy and in addition $z = -\frac{2}{3q}$ describes stiff matter. The second set is

\begin{align}
p= -\frac{1}{3} q z + \frac{1}{3}, \ \ \ \ \lambda_{1}= -\frac{3 \lambda_{2} z^{2}}{2 (qz-1)(2q z +1)}, \ \ \ \ \lambda_{3}= - \frac{3 \lambda_{4}z }{8(q z-1)}, \ \ \ \ \lambda_{6}= - \frac{2\lambda_{4}(qz-1)^{2}}{9 z^{2}}.
\end{align}
Here $\omega_{eff} =  \frac{1+ qz}{1- qz}$,  meaning that the matter dominated era is obtained for $z = -\frac{1}{q}$, radiation dominated epoch for $z = - \frac{1}{2q}$ and de Sitter dark energy for $qz \rightarrow \infty$, with $q=0$ describing the stiff matter. In addition, for $q z > 1$ the model describes phantom dark energy and for $q z < -2$ explains quintessence dark energy. In fact, the model has accelerating phase for $qz>1$ and $qz<-2$ and goes through a decelerating phase for $-2<qz<1$. In the presence of matter density it gives the following solutions

\begin{align}
p =-\frac{1}{3} qz+\frac{2}{3},  \ \lambda_{1} = \frac{3 \rho_{0m}}{a_{0}^{3}\phi_{0}^{q} (qz+4)}, \ \lambda_{2} = \frac{- 2 q (2 q z -1) \rho_{0m}}{a_{0}^{3}\phi_{0}^{q} z},  \ \lambda_{3} = \frac{27 z^{3} \lambda_{6}}{16 (q^{3}z^{3} - 6 q^{2}z^{2}+12q z -8)}, \ \lambda_{4} = -\frac{9 z^{2}\lambda_{6}}{q^{2}z^{2}-4 q z+4},
\end{align}
where we have used Eq. (\ref{dd1}).  This solution has $\omega_{eff}= \frac{qz}{2-qz}$ which describes quintessence dark energy for $q z <-1$, phantom dark energy for $q z> 2$, radiation dominated era for $q z = \frac{1}{2}$ and as we have previously mentioned,  a matter dominated epoch for $q =0$ in the presence of matter density.  To put it in a different way, the universe goes through an accelerating phase for $qz<-1$ and $qz>2$ and experiences a decelerating phase for $-1<qz<2$. In addition, it describes stiff matter for $z = \frac{1}{q}$.

Inserting $q = 0$ in the first and second Friedmann Eqs. (\ref{nb3}, \ref{nb4}) and in the absence of Guss-Bonnet and Galileon term ($\lambda_{3}= \lambda_{5} = 0$), one looks for a solution of the form

\begin{align}
H= \frac{p}{t}, \ \ \ \ \ \ \ \Phi= z.
\end{align}
Substituting the above solution in Eqs. (\ref{nb3}, \ref{nb4}) and utilizing constraint equations coming from the coefficients of $\frac{1}{t^{2}}$ and the rest of the terms, one obtains the following relations between parameters of the model

\begin{align}
p =\frac{2}{3}, \ \ \ \ \lambda_{4} =- \frac{2}{3 z^{2}}, \ \ \ \ \lambda_{6} = \frac{\lambda_{2} z^{-2m +2}}{ 2^{-m+1} m- 2^{-m}+ 2^{-2m}}, \ \ \ \
\lambda_{7}&=- \frac{\lambda_{2} z^{2}(2^{-m}- 2 m+1)}{2^{-m+1}+ 4m-2}.
\end{align}
Therefore, this model describes a matter dominated era for the whole range of values of $m$ with constant self-interacting potential. This is interesting  since it explains a matter dominated era which is able to fall into a de Sitter solution to describe late-time acceleration. In the coming sections, we will derive the condition which illustrates that the de Sitter solution is the attractor of the system.

\subsection{Case $m = n+1$}

For $m = n+1$, the coefficients of $X^{n+1}$ and $X^{m}$ in Eq. (\ref{ss2}) merge together and thus Eqs. (\ref{a9}, \ref{a10}) reduce to the following equation

\begin{align}
 -& (2n+2) a^{3} B^{\prime} \xi_{3}^{\prime} - 3 a^{2} A \xi_{3}^{\prime} -  a^{3} B \xi_{3}^{\prime \prime}+a^{3} B \xi_{4}^{\prime} + 3 a^{2} A \xi_{4} + 2 (n+1)  a^{3} B^{\prime} \xi_{4}=0, \label{m2}
\end{align}
in contrast to  Eqs. (\ref{a3} - \ref{a8}, \ref{a11}) which remain the same. In fact, the functional forms for $f(\phi), \omega(\phi), \xi_{1}(\phi), \xi_{2}(\phi)$ and $V(\phi)$ in (\ref{a12} - \ref{sp6}) are at hand and one is also able to determine the functionality of $\xi_{3}(\phi)$ and $\xi_{4}(\phi)$ by utilizing Eqs. (\ref{a8}, \ref{m2}). Solving Eq. (\ref{a8}), $\xi_{3}(\phi)$ is given by

\begin{align}\label{pp1}
\xi_{3}(\phi) = \frac{\lambda_{5}}{[B(\phi)]^{2n+1}} \exp \left(\int\frac{q}{B(\phi)}d \phi\right).
\end{align}
Here we have used $A(a) = -\frac{q}{3}a$ and $B(\phi)$ is an arbitrary function of $\phi$ and is the same as that obtained for the case $m \neq n+1$ from Eqs. (\ref{af1}). Substituting (\ref{pp1}) in (\ref{m2}) and solving  for $\xi_{4}(\phi)$, one has

\begin{align}
\xi_{4}(\phi) =  \frac{\left(2 \lambda_{5} B^{\prime} (\phi) +\lambda_{6}\right)}{[B(\phi)]^{2n+2}}\exp \left(\int \frac{ q}{B(\phi)}d \phi \right).
\end{align}
In fact Noether symmetry makes a link between the functional forms of the entire functions using $B(\phi)$, the same as that for $\lambda_{5}=0$ ($\xi_{3}(\phi)=0$) in the previous section which means that there are several Noether conserved charges, since selecting each functional form  for $B(\phi)$ results in a different Noether vector whereby there are distinct conserved charges. We again intend to search for analytical solutions using these functional forms. Hence, we  substitute these general forms in Eqs. (\ref{w1}, \ref{w2}) and find

\begin{align}
\nonumber &3 \lambda_{1} \left(H^{2} +  q H \Phi\right) - \frac{1}{2}\lambda_{2} \Phi^{2} - 24 \lambda_{3} H^{3} \Phi + \frac{9}{2} \lambda_{4} H^{2} \Phi^{2} - \left(\frac{1}{2}\right)^{n}\lambda_{5} \left(6 n H \Phi^{2n+1} - q \Phi^{2n+2}\right) + \left(\frac{1}{2}\right)^{n+1} \lambda_{6} (2n+1) \Phi^{2n+2}\\& - \lambda_{7} - \rho_{0m} a^{-3} \exp \left(- \int \frac{q d \phi}{B(\phi)}\right) = 0, \label{ff1}\\
\nonumber & \lambda_{1} \left(3 H^{2} + 2 \dot H + 2 q H \Phi  + q^{2} \Phi^{2} +  q \dot \Phi\right) +\frac{1}{2} \lambda_{2} \Phi^{2}-\lambda_{3} \left(16 H \dot H \Phi +16 H^{3} \Phi+ 8 q H^{2}\Phi^{2} +8 H^{2} \dot \Phi\right)  + \frac{1}{2} \lambda_{4} \Phi^{2} \left(3 H^{2} + 2 \dot H\right) \\&+ \lambda_{4} H \Phi \left(q \Phi^{2} + 2\dot \Phi\right) - \left(\frac{1}{2}\right)^{n}\lambda_{5} \left(2n \Phi^{2n}  \dot \Phi+ q \Phi^{2n+2}\right) - \left(\frac{1}{2}\right)^{n+1} \lambda_{6} \Phi^{2n+2} - \lambda_{7} = 0, \label{ff2}
\end{align}
where in the course of derivation we have divided the whole expression by $\exp \left(\int \frac{q d \phi}{B(\phi)}\right)$ and defined $\Phi = \frac{\dot \phi}{B(\phi)}$. Having a look at Eqs. (\ref{ff1}, \ref{ff2}), one  observes that they contain two variables $H$ and $\Phi$ and that their first order derivatives are similar to the case $m \neq n+1$. Consequently, one may demand a variety of solutions and find conditions on the parameters of the model as was done in the previous section. First, we search for a de Sitter solution by considering

\begin{align}
H &= H_{dS}\Rightarrow a(t) = a_{0} \exp (H_{dS} t), \\ \Phi &= \Phi_{dS}\Rightarrow \int \frac{d \phi}{B(\phi)} = \Phi_{dS} t.
\end{align}
Here we have chosen a zero constant of integration in the second equation without loss of generality with $H_{dS}$ and $\Phi_{dS}$ being constant. Plugging the above solution in Eqs. (\ref{ff1}, \ref{ff2}), for $\rho_{0m}=0$, one finds two constraint equations which enable us to find $H_{dS}$ and $\Phi_{dS}$ in terms of the parameters of the model. Therefore, the model at hand has a de Sitter solution for the selected functional forms via Noether symmetry. Second, we attempt to find a power-law solution for Eqs. (\ref{ff1}, \ref{ff2}), assuming
\begin{align}
H &= \frac{p}{t} \Rightarrow a(t) = a_{0} t^{p},\\  \Phi &= \frac{z}{t} \Rightarrow \int \frac{d \phi}{B(\phi)} = z \ln t +  \ln \phi_{0},
\end{align}
where $\ln \phi_{0}$ is an integration constant. In fact, the evolution of the scalar field is obtained from an integral equation for $\phi$. Therefore, inserting the above solution in Eqs. (\ref{ff1}, \ref{ff2}) and equating the coefficients of $\frac{1}{t^{2}}, \frac{1}{t^{4}}$ and $\frac{1}{t^{2n+1}}$  to zero, one obtains the following results

\begin{align}
p= 2n+1, \ \ \ \ z = - \frac{2(2n+1)}{q}, \ \ \ \ \lambda_{1}= - \frac{2 \lambda_{2}}{3 q^{2}},\ \ \ \
\lambda_{5}= - \frac{\lambda_{6} (2n+1) }{2q(3n+1)},
\end{align}
where $ \lambda_{3}=\lambda_{4}=\lambda_{7}=0$. This means that there is no self-interaction potential and evolution of the universe is controlled by the kinetic term rather than the potential term. There is one solution for which $n=1$ but we exclude that since it will lead to different constraint equations, hence, this case will be investigated in the next section as a special case. The solution explains quintessence dark energy for $ n>0$ but it is not able to describe matter dominated, radiation dominated or phantom dark energy. One can also find the following solution in the presence of matter density

\begin{align}
p= - \frac{1}{3} q z + \frac{2}{3} , \ \ \ \
n= \frac{1}{2}, \ \ \ \
\lambda_{1}= \frac{3 \rho_{0m}}{a_{0}^{3} \phi_{0}^{q} (q z +4)},\ \ \ \
\lambda_{2} = - \frac{2 q (2 q z-1) \rho_{0m}}{a_{0}^{3} \phi_{0}^{q} z (q z+4)}, \ \ \ \
\lambda_{5}= - \frac{\lambda_{6} z}{q z -1},
\end{align}
where we have utilized Eq. (\ref{dd1}). This solution has $\omega_{eff}= \frac{qz}{2-qz}$ which describes quintessence dark energy for $q z <-1$, phantom dark energy for $q z> 2$, radiation dominated era for $q z = \frac{1}{2}$, stiff matters for $z = \frac{1}{q}$ and as we previously mentioned  a matter dominated epoch appears for $q =0$ in the presence of matter density. In other words, the universe goes through an accelerating phase for $qz<-1$ and $qz>2$ and experiences a decelerating phase for $-1<qz<2$.

\subsection{General Brans-Dicke with string correction ($n=1$, $m=2$)}

In this case, the functional forms  found in the case $m=n+1$ are still valid, although the partial differential Eqs. (\ref{a3} - \ref{a11}) change a little. In fact, in Eq. (\ref{ss2}) the coefficients of the term $ \dot a \dot \phi^{3}$ should be added to the coefficients of the term $\dot a \dot \phi X^{n}$, since we assume $n=1$ and the condition $A_{,\phi}=0$ will not consequently exist in this case. Therefore, $A(a, \phi)$ can have the functionality of both $a$ and $\phi$ but as one may deduce by looking at the system of partial differential equations the only possibility to find solutions is when we have $A(a, \phi) \propto a$ and $B(a, \phi) = B(\phi)$ similar to the cases where $m\neq n+1$ and $m=n+1$. However, the exact solutions for scale factor and scalar field change since the constraint equations change. Therefore, putting $n=1$ into equations (\ref{ff1}, \ref{ff2}), one may find two Friedmann equations in terms of $H$ and $\Phi$ for the general Brans-Dicke theory with string correction for which $\Phi = \frac{\dot \phi}{B(\phi)}$. In addition, a de Sitter solution will also exist for this case since this is a subclass of the case $m=n+1$.

Demanding a power-law solution ($H=\frac{p}{t}$ and $\Phi = \frac{z}{t}$) and equating the coefficients of $\frac{1}{t^{2}}$ and $\frac{1}{t^{4}}$ to zero, for $\rho_{0m}=0$, one finds two solutions, the first of which is

\begin{align}\label{sp1}
p= - \frac{1}{2} q z, \ \  \ \
\lambda_{2}  = - \frac{3}{2} \lambda_{1} q^{2}, \ \ \ \
\lambda_{5} = - q (6 \lambda_{3} q + \frac{3}{2} \lambda_{4}), \ \ \ \
\lambda_{6}= \frac{1}{2} q^{2} (24 \lambda_{3} q +  5 \lambda_{4}),
\end{align}
where $\omega _{eff} =-1 - \frac{4}{3qz}$, which means that the universe goes through an accelerating epoch for $qz<-2$ and $qz>0$ and a decelerating epoch for $-2<qz<0$. In other words, the model explains quintessence dark energy for $q z<-2$ and phantom dark energy for $q z>0$. Furthermore, $z = -\frac{4}{3q}$ to  have matter dominated era, $z = -\frac{1}{q}$ to have radiation dominated era  and $qz \rightarrow \infty$ for de Sitter dark energy and moreover, it describes stiff matter for $z = -\frac{2}{3q}$. The second set of solutions result in the following relations between parameters of the model

\begin{align}\label{sp2}
\nonumber p&= - \frac{1}{3} q z + \frac{1}{3},\ \ \ \
\lambda_{2}  =- \frac{2}{3} \frac{(2 q^{2} z^{2} - q z -1) \lambda_{1}}{z^{2}}, \ \ \ \
\lambda_{5} = - \frac{1}{3} \frac{8 \lambda_{3} q^{2} z^{2} + 3 \lambda_{4} q z^{2} -  16 \lambda_{3} q z- 3 \lambda_{4} z + 8\lambda_{3}}{z^{2}},\\
 \lambda_{6}  &=\frac{2}{27 z^{3}}\left(56 \lambda_{3} q^{3} z^{3} + 18 \lambda_{4} q^{2} z^{3} - 144 \lambda_{3} q^{2} z^{2} - 27 \lambda_{4} q z^{2} + 120 \lambda_{3} q z + 9 \lambda_{4} z - 32 \lambda_{3}\right).
\end{align}
Here $\omega_{eff} =  \frac{1+ qz}{1- qz}$  which means matter dominated era occurs for $z = -\frac{1}{q}$,  radiation dominated epoch for $z = - \frac{1}{2q}$ and de Sitter space-time for $qz \rightarrow \infty$. In addition, the model describes phantom dark energy for $q z > 1$ and  explains quintessence dark energy for $q z < -2$. In fact, the model has an accelerating phase for $qz>1$ and $qz<-2$ and goes through a decelerating phase for $-2<qz<1$. Furthermore, it describes stiff matters for $q=0$.

Having a look at Eqs. (\ref{sp1}, \ref{sp2}) and their associated constraint equations, one can deduce that there are several degrees of freedom  and yet because of the existence of string corrections in the action it would be interesting to take one step further and find out if such a model is cable of describing the early, intermediate and late time universe simultaneously. To this end, we consider two power-law solutions, $(p_{1} =\frac{2}{3}$, $z_{1})$ to describe matter dominated epoch and  $(p_{2}= \frac{1}{2}$, $z_{2})$ to explain radiation dominated era which should satisfy two constraint equations coming from the de Sitter solution in order to portray the current accelerating phase. Therefore, we have 8 equations which should  be satisfied simultaneously to have a unified theory describing  radiation and matter dominated era. There are two classes of solutions given by

\begin{align}\label{sr1}
z_{1} = - \frac{4}{3q}, \ \ \ \ z_{2} = - \frac{1}{q}, \ \ \ \ \lambda_{2} = - \frac{3}{2} \lambda_{1} q^{2}, \ \ \ \ \lambda_{5} &= -\frac{3}{2} q(4 \lambda_{3} q + \lambda_{4}),  \ \ \ \ \lambda_{6} = q^{2} (12 \lambda_{3} q + \frac{5}{2} \lambda_{4}),
\end{align}
and

\begin{align}
z_{1}= - \frac{1}{q}, \ \ \ \  z_{2}= - \frac{1}{2 q}, \ \ \ \  \lambda_{2} = 0,\  \ \ \ \lambda_{5} = -3 q (8 \lambda_{3} q+  \lambda_{4}),  \ \ \ \ \lambda_{6} = 8 q^{2} (\lambda_{3} q + \lambda_{4}),
\end{align}
where $\lambda_{1}$ is an arbitrary constant. Putting the above solutions in two constraint equations for the de Sitter solution results in finding $H_{dS}$ and $\Phi_{dS}$ in  terms of  the rest of the parameters of the model. On the other hand, one may demand another power-law solution which has $p> 1$ in order to describe late-time acceleration. As an illustration, taking $p_{3}= 2, z_{3}$ and simultaneously solving the resulting 4 constraint equations with 8 equation signifying radiation and matter dominated era would give us (\ref{sr1}) and $z_{3} = - \frac{1}{4q}$.  Furthermore, it should be stressed that for $q=-1$ and $B(\phi)=1$, one obtains $f(\phi) \sim \omega(\phi) \sim \xi_{i}(\phi) \sim e^{-\phi}$ and $V=0$, where $i$ runs from $1$ to $4$ which is the pre-Big Bang (PBB) scenario of kinetically driven string inflationary cosmology. The cosmological perturbations of such a model have been investigated in \cite{string perturbations} while the authors in \cite{graceful exit} have shown that the model has a graceful exit for some range of parameters of the model.  It should also be emphasized that the model at hand is closely related to the model investigated in \cite{string dark energy} where the authors have considered  $\alpha^{\prime}$ at the tree-level in order to justify its quintessential effects and have shown that it may consistently describe the universe, see also \cite{dark energy gasperini}.

We therefore find that a general Brans-Dicke with string correction and without a self interacting potential of which PBB is a subclass, is able to describe a sequence from inflationary to late-time acceleration  and also has the flexibility to be tested against observational data due to the existence of four free parameters in the model.
In the case of minimal models ($q=0,f=1, \lambda_{1}=1$), the first solution, Eqs.(\ref{sp1}), reduces to $p=0$ for Einstein-Hilbert action which is a trivial solution. The second solution, Eqs. (\ref{sp2}), was found for non-minimal case and reduces to $p = \frac{1}{3}$ which describes a decelerating phase.

One can also find power-law solutions in the presence of matter density as follows

\begin{align}
p& = - \frac{1}{3} q z + \frac{2}{3},  \ \ \
\lambda_{1}= \frac{3 \rho_{0m}}{a_{0}^{3} \phi_{0}^{q}(q z+4)}, \ \ \
\lambda_{2}= - \frac{2 \rho_{0m} q(2 q z-1)}{a_{0}^{3} \phi_{0}^{q} z (q z+4)},\ \ \
\lambda_{5}= -\frac{1}{3 z^{2}} \left(8 \lambda_{3} q^{2} z^{2}+ 3 \lambda_{4} q z^{2}- 32 \lambda_{3} q z - 6 \lambda_{4} z\right.\\&\left.+ 32 \lambda_{3}\right), \ \ \ \
\nonumber \lambda_{6}= \frac{4}{27 z^{3}} \left(28 \lambda_{3} q^{3} z^{3}+ 9 \lambda_{4} q^{2} z^{3} - 144 \lambda_{3} q^{2} z^{2}- 27 \lambda_{4} q z^{2} + 240 \lambda_{3} q z+ 18 \lambda_{4} z - 128 \lambda_{3}\right),
\end{align}
where we have used Eq. (\ref{dd1}). Therefore, the solution has $\omega_{eff} = \frac{qz}{2-qz}$ describing an accelerating phase for  $q z<-1$ and $q z > 2$ and going through a decelerating phase for $-1<qz<2$. In other words, the model describes quintessence dark energy for $q z<-1$, phantom dark energy for $q z > 2$ and radiation dominated era for $q z =\frac{1}{2}$. As was shown before a matter dominated solution in the presence of $\rho_{0m}$ was obtained for $q=0$, indicating a minimal model $f=1$, $\lambda_{1}=1$. Furthermore, it describes stiff matters for $qz=1$.
In addition, one can search for a bouncing universe solution by requesting

\begin{align}
H = b t , \ \ \ \ \  \ \ \  \Phi = z t,
\end{align}
and inserting them into the first and second Friedmann Eqs. (\ref{ff1}, \ref{ff2}). One then finds a solution in the absence of $\rho_{0m}$ which satisfies the constraint equations coming from the coefficients of $t^{2}$ and $t^{4}$ as follows

\begin{align}
b= - \frac{q z}{2}, \ \ \ \
\lambda_{1}= - \frac{2 \lambda_{2}}{3 q^{2}}, \ \ \ \
\lambda_{5} =- q( 6 \lambda_{3} q+ \frac{3}{2} \lambda_{4}), \ \ \ \
\lambda_{6} = \frac{1}{2} q^{2} (24 \lambda_{3} q +  5 \lambda_{4}),
\end{align}
It should be stressed that in the presence of matter one cannot achieve a bouncing universe solution. Therefore, we have a bouncing universe solution as expected since such string corrections usually lead to a non-singular cosmology and,  as is well known, a bouncing cosmology also provides us with  such a non-singular scenario, for a good review on bouncing cosmology see \cite{bouncing cosmology}.

Before closing the section, we summarize our results in Table I by presenting the possible effective equation of state in a general Scalar-Tensor theory utilizing Noether symmetry  and its associated properties. In fact, we have found several models with distinct functional forms  in the presence and absence of matter content $\rho_{0m}$ which have brought about six possible effective equations of state for which the Noether symmetry exists.

\begin{table}[H]
\begin{center}
\begin{tabular}{ |c|c|c|c|c|c|c|  }
\hline
\multirow{2}{5.5em}{EoS for power-law solutions}&\multirow{2}{5.5em}{phantom dark energy}&\multirow{2}{5.5em}{quintessence dark energy}&\multirow{2}{5.5em}{radiation dominated}&\multirow{2}{4.5em}{matter dominated}&\multirow{2}{4.5em}{stiff matters}\\
&&&&&\\
\hline
$\omega_{eff} = - \frac{6n+1}{6n+3}$&$-$&$ n>0$&$-$&$-$&$-$ \\
\cline{1-6}
$\omega _{eff} =-1 - \frac{4}{3qz}$&$q z>0$&$q z<-2$&$q z=- 1$&$q z = -\frac{4}{3}$&$qz = -\frac{2}{3}$ \\
\cline{1-6}
$\omega_{eff} =  \frac{1+ qz}{1- qz}$&$q z>1$&$q z<-2$&$q z = -\frac12$&$q z = -1$&$q=0$\\
\cline{1-6}
$\omega_{eff} = 0$&$-$&$-$&$-$&$for \  all \ m$&$-$\\
\cline{1-6}
$\omega_{eff} = \frac{qz}{2-qz}, \rho_{0m}\neq 0$&$q z> 2$&$qz<-1$&$q z = \frac{1}{2}$&$q =0$&$qz=1$\\
\cline{1-6}
$\omega_{eff} = \frac{q}{q-2}, \rho_{0m}\neq 0$&$1<q<2$&$\frac{1}{2}<q<1$&$q = -1$&$q =0$&$-$\\
\hline
\end{tabular}
\end{center}
\caption{The results in nutshell}
\end{table}

                                                                            \section{Stability of de Sitter solution}

 In this section, the action (\ref{aa1}) is expanded to second order in  perturbation around the dS background where there exists only one propagating scalar degree of freedom which corresponds to the curvature perturbation $\mathcal{R}$. In terms of the gauge invariant quantity $\mathcal{R}$, the second-order perturbed action about the dS background is given by (see details in \cite{inflation-review, second order action})

\begin{align}\label{as1}
 S_{2} = \int d t d^{3} x a^{3} Q_{s} \left[ \dot{\mathcal{R}}^{2}- \frac{c_{s}^{2}}{a^{2}} \left(\nabla \mathcal{R}\right)^{2}\right],
\end{align}
where $Q_{s} $ and $c_{s}^{2}$ are given by

\begin{align}\label{as2}
Q_{s} &\equiv \frac{m_{1} (4 m_{1} m_{3} + 9 m_{2}^{2})}{3 m_{2}^{2}}, \ \ \ \
c^{2}_{s} \equiv \frac{3 (2 m_{1}^{2} m_{2} H - m_{2}^{2} m_{4} + 4 m_{1} \dot m_{1} m_{2} - 2 m_{1}^{2} \dot m_{2})}{m_{1} (4 m_{1} m_{3} + 9 m_{2}^{2})},
\end{align}
and

\begin{align}
m_{1} &\equiv f - 8 H \dot \xi_{1}- \frac{1}{2}\xi_{2} \dot \phi^{2},\\
m_{2} &\equiv \dot f+2 H f  - 24 H^{2} \dot \xi_{1} - 3 H \dot \phi^{2} \xi_{2} - 2 n \dot \phi \xi_{3} X^{n},\\
m_{3} &\equiv - 9 H^{2} f - 9 H \dot f +3 \omega X + 144 H^{3} \dot \xi_{1} + 27 H^{2} \dot \phi^{2} \xi_{2} + 18 n(n+1) H \dot \phi \xi_{3} X^{n}- 6(n+1) \xi_{3}^{\prime} X^{n+1}
- 3 m (2m-1) \xi_{4} X^{m},\\
m_{4} &\equiv f - 8 \ddot \xi_{1} +\frac{1}{2} \dot \phi^{2}\xi_{2}.
\end{align}
In order to avoid the appearance of ghosts and Laplacian instabilities in the theory we require that $Q_{s}> 0$ and $c_{s}^{2}>0$, respectively. One can  calculate $Q_{s}$ for the functions  found in the case $m=n+1$ as follows

\begin{align}
m_{1} &= \left(\lambda_{1} - 8 \lambda_{3} H_{dS} \Phi_{dS} - \frac{1}{2} \lambda_{4}\Phi_{dS}^{2}\right) \exp \left(\int \frac{q}{B(\phi)}d \phi\right),\\
m_{2} &= \left(q \lambda_{1} \Phi_{dS}+ 2\lambda_{1} H_{dS} - 24 \lambda_{3} H^{2}_{dS} \Phi_{dS} - 3\lambda_{4} H_{dS} \Phi^{2}_{dS}- 2n \left(\frac{1}{2}\right)^{n} \lambda_{5}\Phi^{2n+1}_{dS}\right) \exp \left(\int\frac{q}{B(\phi)} d \phi\right),\\
\nonumber m_{3}&= \left(-9\lambda_{1} H^{2}_{dS} - 9 \lambda_{1} q H_{dS} \Phi_{dS} + \frac{3}{2} \lambda_{2} \Phi^{2}_{dS} + 144 \lambda_{3} H^{3}_{dS} \Phi_{dS} +27\lambda_{4} H^{2}_{dS} \Phi^{2}_{dS} + 18 n (n+1) \left(\frac{1}{2}\right)^{n}\lambda_{5}H_{dS} \Phi^{2n+1}_{dS} \right.\\& \left.- 6 (n+1) \left(\frac{1}{2}\right)^{n+1}q\lambda_{5} \Phi^{2n+2}_{dS}- 3 (n+1) (2n+1) \left(\frac{1}{2}\right)^{n+1} \lambda_{6} \Phi^{2n+2}_{dS} \right) \exp\left(\int \frac{q}{B(\phi)} d \phi\right).
\end{align}
Therefore, putting the above expressions in (\ref{as2}), one infers that

\begin{align}
Q_{s} \propto \exp\left(\int \frac{q}{B(\phi)} d \phi\right),
\end{align}
where for de Sitter solution we have

\begin{align}
\frac{\dot \phi}{B(\phi)} = \Phi_{dS} \Rightarrow \int \frac{q}{B(\phi)} d \phi = q \Phi_{dS} t.
\end{align}
Here we have assumed that the constant of integration is zero without loss of generality. One can obtain the Euler-Lagrange equation for action (\ref{as1}) in Fourier space as follows

\begin{align}
\frac{1}{a^{3} Q_{s}} \frac{d}{d t}(a^{3} Q_{s} \dot{\mathcal{R}})+ c^{2}_{s} \frac{k^{2}}{a^{2}} \mathcal{R} = 0,
\end{align}
where $k$ is the co-moving wave number. For homogeneous perturbation ($k=0$) which has only time-dependence, the solution of the above equation is given by

\begin{align}
\mathcal{R}(t) = c_{1} + c_{2} \int \frac{d t}{a^{3} Q_{s}},
\end{align}
where $c_{1}$ and $c_{2}$ are constants of integration. Since the scale factor evolve as $\exp (H_{dS} t)$ and $Q_{s} \propto\exp (q \Phi_{dS} t)$,  the homogeneous perturbation about the dS background evolves as

\begin{align}
\mathcal{R}(t) = c_{1} + \hat{c}_{2} \exp\left(-\left(3 H_{dS} + q \Phi_{dS}\right) t \right),
\end{align}
where $\hat{c}_{2}$ is a constant. In order to avoid the growth of $\mathcal{R}$, one requires that

\begin{align}\label{sd1}
3 H_{dS}+ q \Phi_{dS}>0,
\end{align}
which corresponds to stability condition for the dS solution. This condition will also exist for the rest of cases, although, the functional forms are different.  One can immediately conclude that the matter dominated solutions which were found throughout the paper will fall into de Sitter accelerating phase. Furthermore, the sequence which were found for the case $(n=1, m=2)$ will finally fall into a stable de Sitter phase. Therefore, we finalized our discussion by finding the stability condition (\ref{sd1}) for de Sitter solution.

                                                                                  \section{Concluding remarks}

Throughout the paper, we have discussed  a general method to realize analytical cosmological solutions in general Scalar-Tensor theories. The approach is based on exploring Noether symmetry for a particular dynamical system whereby the dynamical system is reduced and, in fact, allows one to solve the equations of motion. Additionally, such an approach can be contemplated as a physically motivated criterion due to the fact that such symmetries are always associated with conserved quantities.

The prime point is that the existence of Noether symmetry specifies the form of undetermined functions appearing in the action as well as the corresponding point-like cosmological Lagrangian where the FLRW metric is adopted. It deserves stressing that starting from a point-like FLRW Lagrangian and consequently deriving equations of motion  results in the same equations obtained by adopting the FLRW metric in Einstein field equations. This circumstance allows one to explicitly  search for Noether symmetries in the point-like Lagrangian and then to plug the associated conserved quantities into equations of motion. Consequently, the form of the undetermined functions in the action, $f(\phi), \omega(\phi), V(\phi)$ and $\xi_{i}(\phi)$ where $i$ runs from $1$ to $4$ is fixed by demanding the  existence of symmetry conditions. This would immediately simplify the dynamical system as some of its variables (at least one) become cyclic. Therefore, one can utilize cyclic variables and associated conserved charges in order to obtain analytical solutions. In our case, the selected functions symmetrize equations of motion in such a way as to give us the opportunity to find exact solutions by directly solving equations of motion instead of exploiting cyclic variable. In this paper, we have applied Noether symmetry not only to obtain analytical solutions for particular models but also implicitly to identify cosmologically viable models by demanding a sequence extending from an inflationary era to an accelerating phase at late times, as the minimal criteria.

To start with, using Noether symmetry, we found three cases, $m \neq n+1, m=n+1, n=1$ which led to three general forms for the functions appearing in the action. The interesting point regrading the resulting functional forms is that for the case $m \neq n+1$, Noether symmetry just exists for the so-called exponential and power-law coupling. For all three cases, we then attempted to find  possible solutions such as de Sitter, power-law and bouncing  in the case of the vacuum and matter dominated universe and illustrated that there usually exists a deceleration-acceleration transition in Scalar-Tensor theories selected via Noether symmetry. Another interesting point is that the de Sitter solution always exists for the selected functions via Noether symmetry not only for exponential and power-law functions but also for a general form for which the form of functions are determined by the functionality of the second component  of Noether vector $B(\phi)$.  Moreover, we showed that in the presence of matter density, matter dominated solutions exist just for the minimal case ($q=0, \lambda_{1}=1, f=1$). Next, we found a model which is able to show a sequence from an inflationary era to an accelerating epoch at late times. Consequently, we found a general Brans-Dicke with string correction without a self  interacting potential $(n=1,m=2)$ which reduces to the pre-Big Bang scenario (PBB) for $B(\phi)=1$ and $q =-1$ for which the model simultaneously has radiation dominated, matter dominated and de Sitter solutions. In addition, we classified the models based on  parameters involved and the effective equations of state, based on being either phantom or quintessence dark energy for which the results are summarized in Table I. Finally, expanding the action up to second order, we found the condition for stability of de Sitter solution and showed it to be an attractor of the system for $3 H_{dS} + q \Phi_{dS}>0$ in all five cases.
It deserves mentioning at this point that symmetries are not only a mathematical tool to solve dynamical systems but also bring about the opportunity to physically select an observable universe and  ``particularly'' to single out analytical models related to observation \cite{discriminate model} which would provide a tool to classify dark energy models related to Noether symmetry, see \cite{a1}.

As the final remark, there is the question of frame in which an action is considered. Throughout this work, we have adopted the so-called Jordan frame for which the action is given by (\ref{aa1}). As is well known, upon a conformal transformation, the action could in principle be transformed to the so called Einstein frame. This would, however, have taken us too far afield due to the complexity involved. Still, this is an interesting question that deserves to be studied.



\end{document}